\documentclass{article}

\usepackage{arxiv}

\usepackage[utf8]{inputenc} 
\usepackage[T1]{fontenc}    
\usepackage{hyperref}       
\usepackage{url}            
\usepackage{booktabs}       
\usepackage{amsfonts}       
\usepackage{nicefrac}       
\usepackage{microtype}      
\usepackage{lipsum}		
\usepackage{graphicx}
\usepackage{natbib}
\usepackage{doi}

\title{Measuring ancient technological complexity and its cognitive implications using Petri nets}

\author{ \href{https://orcid.org/0000-0002-1216-821X}{\includegraphics[scale=0.06]{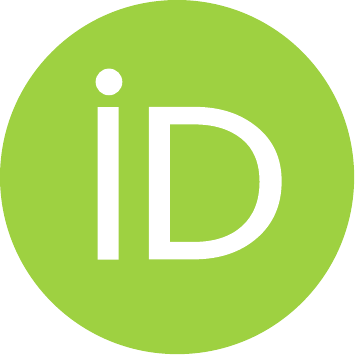}\hspace{1mm}Sebastian Fajardo}\thanks{Corresponding author}\\
	Department of Materials Science and Engineering\\
	Delft University of Technology\\
	Delft, Zuid-Holland, 2628CD, The Netherlands \\
	\texttt{s.d.fajardobernal@tudelft.nl} \\
	\And
	\href{https://orcid.org/0000-0000-0000-0000}{\includegraphics[scale=0.06]{orcid.pdf}\hspace{1mm}Paul R. B. Kozowyk} \\
	Department of Materials Science and Engineering\\
	Delft University of Technology\\
	Delft, Zuid-Holland, 2628CD, The Netherlands \\
	\texttt{P.R.B.Kozowyk@tudelft.nl}\\
	\AND
	\href{https://orcid.org/0000-0003-3640-9054}{\includegraphics[scale=0.06]{orcid.pdf}\hspace{1mm}Geeske H. J. Langejans} \\
	Department of Materials Science and Engineering / Palaeo-Research Institute\\
	Delft University of Technology / University of Johannesburg\\
	Delft, Zuid-Holland, 2628CD, the Netherlands / Johannesburg, Gauteng 2092, South Africa \\
	\texttt{g.langejans@tudelft.nl}\\
}

\renewcommand{\headeright}{}
\renewcommand{\undertitle}{}
\renewcommand{\shorttitle}{Measuring ancient technological complexity}

\hypersetup{
pdftitle={A template for the arxiv style},
pdfsubject={q-bio.NC, q-bio.QM},
pdfauthor={David S.~Hippocampus, Elias D.~Striatum},
pdfkeywords={First keyword, Second keyword, More},
}

\begin{document}
\maketitle

\begin{abstract}

We implement a method from computer sciences to address a challenge in Paleolithic archaeology: how to infer cognition differences from material culture. Archaeological material culture is linked to cognition: more complex ancient technologies are assumed to have required complex cognition. We present an application of Petri net analysis to compare Neanderthal tar production technologies and tie the results to cognitive requirements. We applied three complexity metrics, each relying on their own unique definitions of complexity, to the modelled production sequences. Based on the results, we suggest that Neanderthal working memory requirements may have been similar to human preferences regarding working memory use today. This method also enables us to distinguish the high-order cognitive functions combining traits like planning, inhibitory control, and learnings that were likely required by different ancient technological processes. The Petri net approach can contribute to our understanding of technology and cognitive evolution as it can be used on different materials and technologies, across time and species.
\end{abstract}


\section{Introduction}

Human origins and the evolution of cognition are intricately tied to the use of technology (\cite{lombardCognitionCapuchinRock2019, nowellStoneToolsEvolution2010, overmannSqueezingMindsStones2019, roebroeksNeandertalsRevised2016, wadleyCompoundadhesiveManufactureBehavioral2010, wynnExpertNeandertalMind2004}). The development of complex technologies over the last 3.3 million years provides a mirror to the cognitive developments that underpin behavioral changes. Generally, the processes of production and use of archaeological objects are first (modelled and/or experimentally) reconstructed and then interpreted using concepts such as cognitive load, learning, reflectiveness, working memory, extended thought, and action sequences (\cite{goldenbergNeuralBasisTool2009, hodgsonSymmetryAcheuleanHandaxes2015, lombardThinkingBowandarrowSet2012, stoutLateAcheuleanTechnology2014}). Oversimplifying one of the main hypotheses, it could be said that a more complex mind can give rise to more complex technologies and thus that we can reverse engineer cognition from technology and material culture. However, the link between the complexity of technologies and cognition remains qualitative, restricting systematic comparisons of different technological behaviors and their cognitive requirements. 

Tar production, an example of complex technology, often features in discussions about Neanderthal and modern human technological and cognitive capabilities (\cite{kollerHightechMiddlePalaeolithic2001, niekusMiddlePaleolithicComplex2019, roebroeksNeandertalsRevised2016, schmidtArchaeologicalAdhesivesMade2022}). However, the exact complexity of birch tar technology is debated, as there are multiple ways to make tar without fireproof containers (\cite{kozowykReplySchmidtInterpretation2020, schmidtBirchTarProduction2019, schmidtSteakTournedosBeef2021}). Recent experiments show that birch tar can be produced with simple methods (\cite{schmidtBirchTarProduction2019}). However, none of the reconstructed methods have been systematically studied for their complexity with definitions for what is considered simple or complex that can contribute to the current technology and cognition debates. Condensation, the simplest method, does require less materials and the production process consists of fewer unique steps than other techniques, but the implications of these criteria on complexity/cognition are unspecified. In this paper we take a step back in the debate and we explore a method to overcome these two problems of a) measuring technological complexity, and b) linking technology to cognition. We use Petri net modelling (\cite{fajardoModellingMeasuringComplexity2022}) to compare the complexity of Neanderthal birch tar production methods in terms of the cognitive requirements of their technological behaviors. 

The measurement and comparison of different technological processes is often challenging due to the uniqueness of interrelations between cognitive processes and technological behaviors. Various measures have been used in the past, such as counting techno-units in tool kits, steps in behavioral sequences, procedural units in the production of a specific tool, number of and/or decisions, and distance between a need and the satisfaction of that need (\cite{klinePopulationSizePredicts2010, kozowykExperimentalMethodsPalaeolithic2017, lombardCognitionCapuchinRock2019, mullerMeasuringBehaviouralCognitive2017, oswaltTechnologicalComplexityPolar1987, perreaultMeasuringComplexityLithic2013}). These measurements are generally unique for specific tools and materials. In this paper we use Petri nets as a tool for expanding measurements of technological processes. Petri nets can model and study the causality and execution of events, including sequential, concurrent and parallel execution (\cite{fajardoModellingMeasuringComplexity2022}). This allows the identification of differences in the way information is processed to obtain or use a product. Additionally, we argue that with Petri nets, different definitions for complexity can be applied and compared for the same production process, exposing related behavioral and cognitive implications. Our approach moves away from focusing solely on verifying one particular dimension (definition) of technological complexity. Instead, we consider the differences between production methods in light of the type of solution a method represents. 

Previous studies suggest that the production of Paleolithic or Stone Age adhesives requires the maker to be able to: a) accessing multiple pieces of information at the same time while executing the process; b) avoid errors and correct problems throughout the production process; c) understand and abstract information about the materials, product templates, and the process itself before starting to make adhesives (\cite{lombardFourfieldCoevolutionaryModel2021, wadleyCompoundadhesiveManufactureBehavioral2010, wynnExpertCognitionModel2016}). 
Accessing information at the same time (a) is closely related to working memory, which is one key cognitive feature that extends over the evolution of the human mind (\cite{coolidgeWorkingMemoryIts2005}). The requirements of working memory can be identified by the number of interactions between elements in a process; the more interaction, the higher the cognitive load (\cite{figlCognitiveComplexityBusiness2011, wangMeasureMentalWorkload2020}). Further, the reliance on working memory and higher cognitive functions (\cite{raduntzEffectPlanningStrategy2020}) can be identified as the number of features retained in memory during natural behavior (\cite{draschkowWhenNaturalBehavior2021}). Conscious attention and working memory also appear to be closely linked. For example, working memory serves to focus attention and make decisions (\cite{coolidgeRiseHomoSapiens2009}). An important component in working memory, described as ‘executive attention’ (\cite{kaneRolePrefrontalCortex2002}), also functions to maintain neural stimuli that are relevant to reaching the end goal of a task in the face of interference-rich contexts (\cite{coolidgeRiseHomoSapiens2009}). 

Multiple permutations of events increase the likelihood of errors (b) and complexity. In a production process, different permutations create more possible paths to obtain the final product and increase the probability of deadlocks, wasting time and resources (\cite{parkEmpiricalEstimationHuman2021}). Complexity is also increased with more choices, because individuals may be unaware of all available choices, or what the best choices are to obtain the desired product. For these reasons, a sequential production process is also easier to execute than one with several paths to reach the end of the workflow  (\cite{ranganathanWhatComplexityDistributed2007}). Implementing strategic planning (\cite{fragaszyStrategicNavigationTwodimensional2003}) and inhibitory control (\cite{shenoyRationalDecisionmakingInhibitory2011}) may reduce the likelihood of errors produced by different permutations and choices and help to solve complex problems. 

Process structures with more elements and relations are harder to understand (c) because more information needs to be processed. Since information presented in a simple arrangement is easier to understand than if the same information is presented in an elaborate structure, both the amount of information and the types of structures affect the structural complexity of a process. This in turn affects process understandability, which provides an indication of how much information is embedded in the process (\cite{dikiciFactorsInfluencingUnderstandability2018}). Previous studies have identified that understanding how to produce a technology was an important aspect in the acquisition, transmission, and production of Paleolithic technologies (see for example \cite{nonakaHowStoneKnappers2010, stoutLateAcheuleanTechnology2014}; but see also (\cite{lombardHuntingHuntingTechnologies2015, pargeterKnowledgeVsKnowhow2020}).

The complexity of the production of prehistoric artifacts can be measured using the requirements introduced above in combination with computational models (\cite{hoffeckerComplexityNeanderthalTechnology2018}). To do this, we used Petri nets to model Paleolithic tar production processes. Petri nets are a modelling language with underlying mathematical semantics (\cite{fajardoModellingMeasuringComplexity2022, reisigUnderstandingPetriNets2013}). These nets are used to study systems that may show concurrent agents or events and components that operate independently with occasional resource sharing or synchronization. We used workflow nets (\cite{vanderaalstApplicationPetriNets1998}), a class of Petri net, to model and measure the complexity of resulting models using three pre-existing metrics: a) the density metric that takes into account the interconnectedness between events and resources (\cite{mendlingTestingDensityComplexity2016}), and can be related to working memory; b) the extended cyclomatic metric (\cite{lassenComplexityMetricsWorkflow2009}), that concerns the likelihood of errors throughout the process, and the potential need for planning and inhibition control; and c) the structuredness metric, which relates to the effort to understand abstract information about the materials, product templates and the process itself, and thus to learning (\cite{lassenComplexityMetricsWorkflow2009}).

We model and measure three experimental techniques of birch bark tar production known from the literature: condensation (\cite{schmidtBirchTarProduction2019}), pit roll, and raised structure (\cite{kozowykExperimentalMethodsPalaeolithic2017}). Currently, the condensation method is interpreted to be the simplest of the three, and the raised structure the most complex. These methods cover the widest range of potential tar making techniques and represent our current knowledge about aceramic tar production, both in terms of yield, time invested, number of production steps, and materials required. The Petri net models of these production processes and the results of the complexity metrics allow us to present a multidimensional comparison of the complexity of an ancient technology. With these cognitively related metrics, we can also show that, contrary to current ideas (\cite{kochiyamaReconstructingNeanderthalBrain2018, wynnExpertNeandertalMind2004}), working memory, inhibitory control, and planning are all cognitive requirements involved in the aceramic production of tar.

\section{Methods}

Author2, and author3 recreated the tar production processes (\cite{kozowykExperimentalMethodsPalaeolithic2017, schmidtBirchTarProduction2019}) in field experiments. Non-participant observation was implemented to record technical behaviors during the process of each experiment (\cite{czarniawskaFieldworkTechniquesOur2018}). Author1 observed and recorded the experiments and Author2 and Author3 actions with a video camera, and time stamped for all events during the coding phase. After the experiments, Author2 was asked to describe the workflow of the experiments from the executant’s perspective. These different sources were integrated to produce a comprehensive list of the actions, events, conditions and sequences that occurred during the workflow.

\subsection{Petri nets}

Petri nets are directed bipartite graphs with three basic elements: places, transitions, and arcs (e.g. Figure 1). Places in a Petri net represent states, conditions, or resources that need to be met or available before an event can be carried out. They can also represent the result of an event, i.e., a new state or condition of a resource. Places cannot be directly connected with one another.

\begin{figure}
	\centering
	\includegraphics[width=0.8\linewidth]{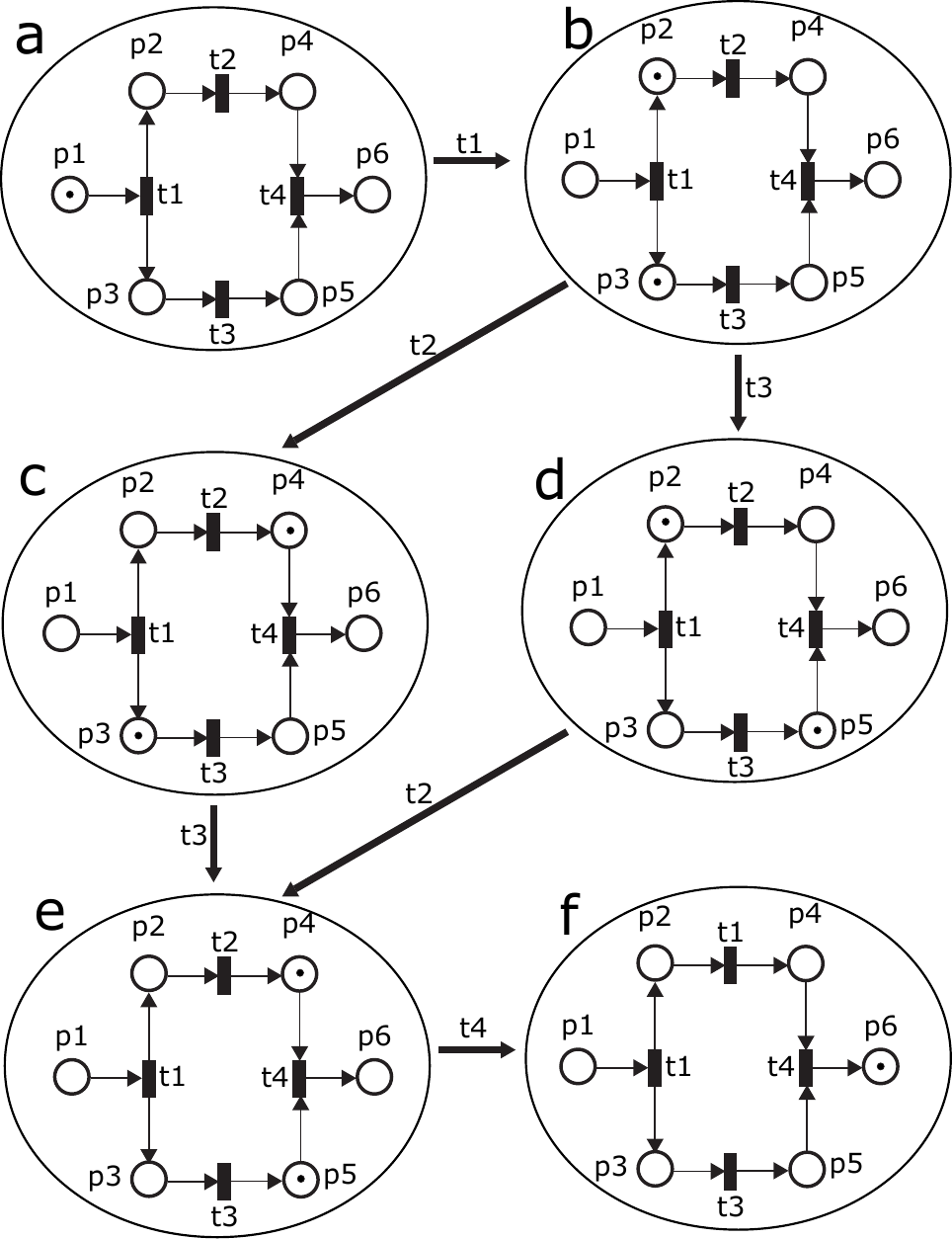}
	\caption{A workflow Petri net representation and reachability graph to illustrate dynamics. The Petri net consists of six places labeled p1 to p6, four transitions labeled t1 to t4, and arcs connecting them. All arcs and transitions have the same function that returns 1, meaning that any transition can fire as long as its input places are marked with tokens. Each subfigure (oval) represents a reachable state (marking) and together with the directed edges (arrows) represent the reachability graph of the Petri net.  In the initial reachable state of the Petri net (a), only place p1 has one token, and all other places are empty. When t1fires, the marking of the Petri net changes and places p2 and p3 have one token each (b). Then transitions t2 and t3 can occur in any order, including in parallel, to produce tokens in p4 and p5, respectively, as shown in (c) and (d). After places p4 and p5 are marked with one token each (e), then transition t4 may occur and produce a token in p6 to reach the final reachable state of the Petri net (f).}
	\label{fig:fig1}
\end{figure}

Transitions in a Petri net represent events that change conditions or states of resources. They can be enabled or disabled depending on the availability of the required resources or the fulfilment of the necessary conditions. Two transitions cannot be directly connected. A transition may occur when all the conditions and resources of its input places are available. The firing of a transition is instantaneous and the choice of which transition to fire when several transitions are enabled at the same time is random. When several transitions are enabled, these transitions may occur concurrently (\cite{fajardoModellingMeasuringComplexity2022, reisigUnderstandingPetriNets2013}).

Arcs in a Petri net represent the relations between places and transitions. They indicate the resources or conditions required for a transition to be enabled or the resources or conditions resulting from a transition. Arcs can be either input arcs, output arcs, or both, depending on their direction and function. By combining these three elements, Petri nets provide a graphical representation of the behavior of a system. It is important to note that Petri nets are not just a graphical representation of a system, but can also be used to analyze the behavior of the system.

In a Petri net the availability of resources or the fulfillment of conditions is graphically represented using black dots or numbers called tokens. They are used to track the flow of information and resources in the system. Tokens are placed in the input places of a transition to indicate that the required resources or conditions are available for the transition to be enabled. When a transition is fired, tokens are removed from the input places and produced in the output places, representing the new state or condition of the resources. The distribution of tokens in different places at any given moment represents the current state of the system and provides insight into the behavior of the system over time. The states of the process are represented by the distribution of tokens over places. These states are also called markings in Petri nets. 

In workflow nets, arcs have a weight of one because places correspond to conditions that can be validated as true or false. Processes modeled as workflow nets start with a token marking one unique input place and should always be able to end with a token in a different unique output place, with all the other places being empty. In workflow nets, transitions can always occur by following the appropriate route in the workflow and they do not create infinite loops (\cite{vanderaalstApplicationPetriNets1998, lassenComplexityMetricsWorkflow2009}).

\subsection{Modelling approach}

The tar production models relied on a set of assumptions. First, we focused on the intrinsic variability of tar production processes, rather than the way environmental settings determine the availability of resources. Therefore, we assumed that cultural, social and environmental restrictions related to resource availability did not play a role in the workings of the processes. Second, we considered that resources, tools, and time required for activities were available and did not represent behavioral constraints. Third, models were developed as action oriented, and people executing actions were excluded from the models. 

For the models presented here, we defined the atomic units as actions or events that changed the location or modified the physical properties of resources. In the tar production models, places represented either the presence/absence of a material, or whether an action occurred. The models ran a single process instance. This enabled the use of pre-existent complexity metrics and controlled the effects of parameters such as the required amount of tar. Assuming infinite resources and time, any of the tar production techniques can be repeated as many times as needed to obtain a specific amount of tar, but these repetitions do not change the workings of each process. For example, for the condensation method, we modelled the events from when one piece of bark was burned, until the tar produced by that piece of bark was stored. However, in practice, this method involved burning several pieces of bark and extracting and storing tar repeatedly in the same way to obtain more tar. 
We restricted the maximum number of tokens that each place can hold to one. This restriction is commonly used in workflow nets to ensure that time complexity was not a practical constraint for the calculation of the metrics (\cite{lassenComplexityMetricsWorkflow2009}). Finally, causal logic determined the control flow of the models. When actions needed to be executed for a certain amount of time, the beginning and end were represented by start-and end-activities, with a place in between denoting ‘in progress’.

The Petri nets were saved as pnml files, a XML-based interchange format for Petri nets, using Snoopy 2 version 1.22 (\cite{heinerSnoopyUnifyingPetri2012}). To analyze the Petri nets models, we calculated the three metrics using ProM Tools release 6.10 developed by the Process Mining Group at Eindhoven Technical University (\cite{vanderaalstProMComprehensiveSupport2007, promProM102010}). We imported pnml files produced in Snoopy to ProM 6.10 using the plug-in PNML Petri net files. We used the plug-in Petri-net Metrics in ProM 6.10 for calculating the metrics. Reachability graphs were also calculated using ProM 6.10.

\subsection{Metrics}

Density metric. The density metric measures the degree of connection between actions and conditions in the process (\cite{mendlingTestingDensityComplexity2016}). It can be used as a proxy for how much information a maker has to access when several actions can be executed at the same time or when several conditions are needed to execute a given action in the process. This metric was originally formulated to characterize networks and later adapted to measure the density of connections in workflow notations, including workflow nets (\cite{mendlingTestingDensityComplexity2016}). The density metric in a workflow net calculates the ratio of existing arcs to the maximum number of possible arcs. The maximum number or arcs is calculated by multiplying the total number of places and transitions and then multiplying the result by two. A high ratio of existing arcs means that conditions and actions are more interconnected, and therefore the amount of information required to change states of the process is higher during behavior. For instance, in Figure 1, the Petri net example has six places and four transitions, which implies that the maximum number of arcs possible is 48. The actual number of arcs present in the Petri net is 10. As a result, the density metric value for this Petri net is calculated to be 0.208.

Extended Cyclomatic metric. The extended cyclomatic metric measures the number of possible paths in which a product can be obtained given the structure of each production method (\cite{lassenComplexityMetricsWorkflow2009}). We use it as a proxy for the likelihood of reassessments and errors throughout the process. The extended cyclomatic metric is calculated with the reachability graph of a Petri net model. A reachability graph calculates the reachable states of a Petri net, that is all of the different moments in which the production system may be observed before obtaining the product (\cite{fajardoModellingMeasuringComplexity2022}). The calculation of the extended cyclomatic metric includes the number of strongly connected components of each reachability graph. A Petri net's reachability graph represents each reachable state of the system being modeled as a vertex. All states (vertices) form the state space of the system. Transitions that occur between states are represented by directed edges (arcs) between vertices. A strongly connected component is a maximal set of reachable states, where each reachable state can be reached from any other state in the component. The minimal strongly connected component in a reachability graph is represented by a single reachable state. The extended cyclomatic metric is calculated by subtracting the total number of reachable states from the total number of directed edges and adding the number of strongly connected components to the result (\cite{lassenComplexityMetricsWorkflow2009}). Practically, this means that low values in the extended cyclomatic metric imply less possible paths to reach the product, and lower chances of producing errors. High values mean that several different paths exist which translates to more possible errors during the process and higher behavioral complexity. In Figure 1, there are six vertices (a,b,c,d,e,f), six directed edges (thick arrows), and six strongly connected components because all reachable states are minimal strongly connected components. As a result, the extended cyclomatic metric value for the Petri net in Figure 1 is six.

Structuredness metric. The structuredness metric measures the effort required to understand the information behind the process (\cite{lassenComplexityMetricsWorkflow2009}). We use the structuredness metric to deconstruct the Petri net models into components similar to programming constructs such as sequences, selections and iterations (\cite{vanderaalstWorkflowPatterns2003, dikiciFactorsInfluencingUnderstandability2018}). These constructs are given a weight based on their structural complexity, which is a reported factor that reduces the comprehension of conceptual models (\cite{lassenComplexityMetricsWorkflow2009, winterMeasuringCognitiveComplexity2020}). In Petri nets, these constructs are sets of places, transitions and arcs. Constructs with a smaller number of elements allow the information about the process to be transmitted easier and more accurately. This makes it easier to store information about components for future use, meaning the process is easier to learn. The algorithm to calculate the metric searches for seven types of components: (1) sequence, (2) choice, (3) while, (4) marked graph, (5) state machine, (6) well-structured, and (7) unstructured (\cite{vanderaalstWorkflowPatterns2003, reisigUnderstandingPetriNets2013}). A sequence is an event that is enabled unconditionally after the completion of another event. A choice is a point in the process where an event of several possibilities is chosen. A while is an event that can be repeated. A marked graph is a process where there are no choices between events, but events can occur concurrently. A state machine is a process that changes states after the occurrence of an event, which can be in conflict with other events, but these and other events cannot occur concurrently. A well-structured component is a process where choice and synchronization between events are separated in the process and there are no cycles. An unstructured component is a minimal process that does not fall within any of the types of components above.
 
The structuredness metric is calculated using a function that weighs each component following the order above. First, the algorithm searches for sequences; the less complex components which receive the lowest weight, and it finishes with unstructured components, which receive the highest weight. Next, a weight is calculated for the identified component which is folded into a single transition. Then, the algorithm continues to search, weight, and fold components in the Petri net using the same priority function. This procedure is conducted until a single transition connected to the initial and final place remains. The weight of the last transition represents all the weights of the identified components. The metric gives higher weights to components containing embedded components, based on the assumption that components made of other nested components are more complicated and therefore harder to understand. A low metric score means a process has low complexity because the amount of information embedded in its structure is small. A high value indicates that the amount of information is large and it will require more effort to understand the structure of the process. For instance, to calculate the structuredness metric of the Petri net in Figure 1, we apply the algorithm by Lassen and van der Aalst (\cite{lassenComplexityMetricsWorkflow2009}). This Petri net is only a marked graph, and its weight function is determined by doubling the number of transitions and multiplying it with the diff function, which measures how evenly split and merge points are matched within the component. A higher diff value indicates unevenness in a workflow net can cause imbalances in the flow and limit the firing of certain transitions that require multiple tokens from different places. The marked graph in Figure 1 has four transition and a diff value of one, making the structuredness metric value equal to eight. Refer to Lassen and van der Aalst (\cite{lassenComplexityMetricsWorkflow2009}) for further details on the weight functions and algorithm.

\section{Results} 

A comparison of the three workflow nets showed that the condensation and pit roll method have less conditions (places), events (transitions) and relations (arcs) than the raised structure method (Figure 2, Figure 3). The condensation model has the smallest number for all elements (Figure 2A, Figure 3). The pit roll model is slightly larger in each element category (Figure 2B, Figure 3), and the raised structure model is the largest in all categories (Figure 2C, Figure 3). In each net, places represent approximately 24\%, transitions 24\%, and arcs 52\% of all elements. These results suggested that the models systematically represent the observed process behaviors with little ambiguity in different types of dependency relations.

\begin{figure}
	\centering
	\includegraphics[width=0.95\linewidth]{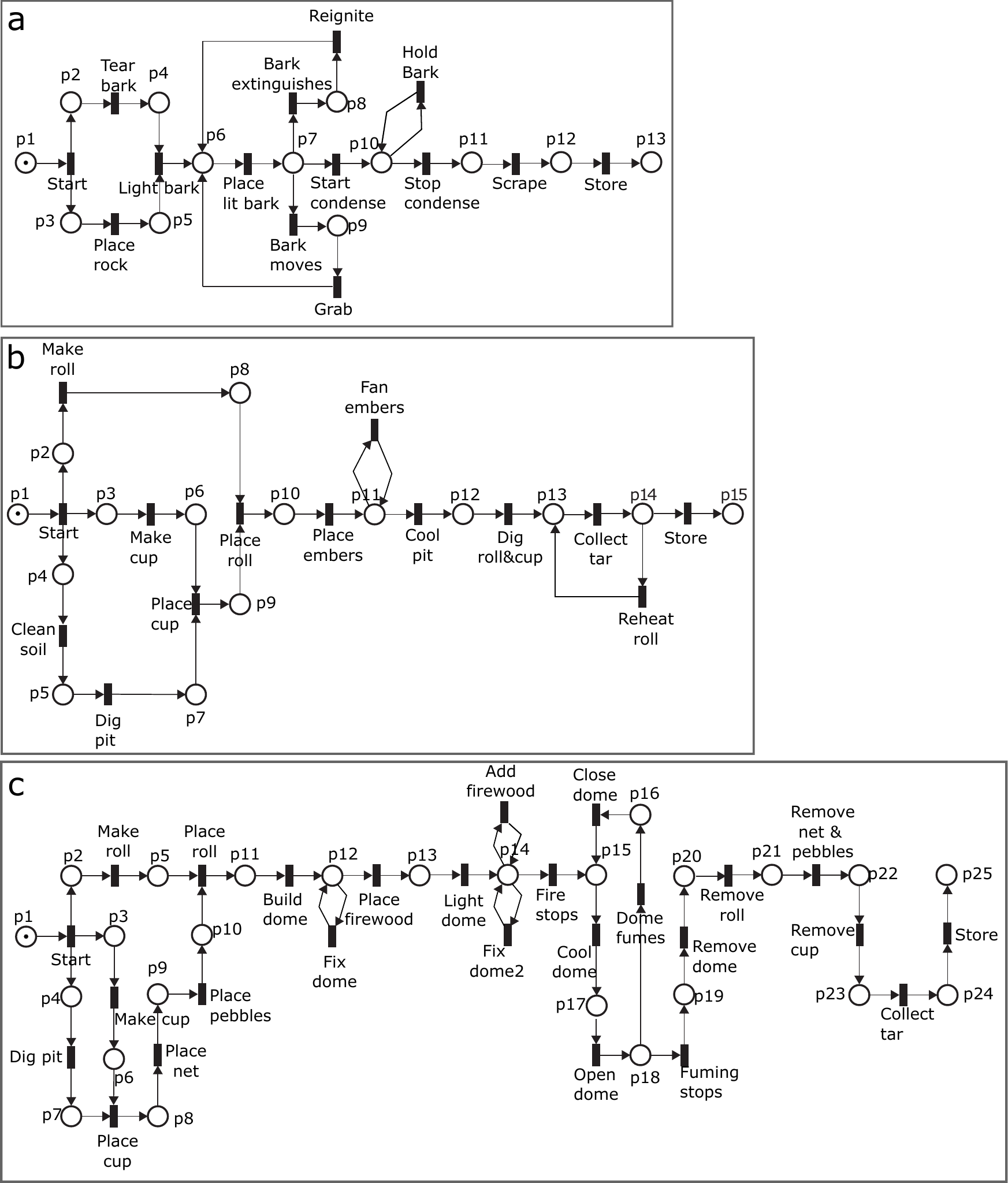}
	\caption{Initial state of Petri net models for the three tar production methods. condensation (a); pit roll (b); raised structure (c). Places (circles) represent the conditions or resources before or after an event. Transitions (rectangles) represent events that change the local states of the system. Arcs (arrows) are directed and form logical connections between places and transitions. They indicate the flow of the system and the causal relations between places and transitions. Token inside places (black dots) represent availability of resources or the fulfillment of conditions}
	\label{fig:fig2}
\end{figure}

\begin{figure}
	\centering
	\includegraphics[width=0.8\linewidth]{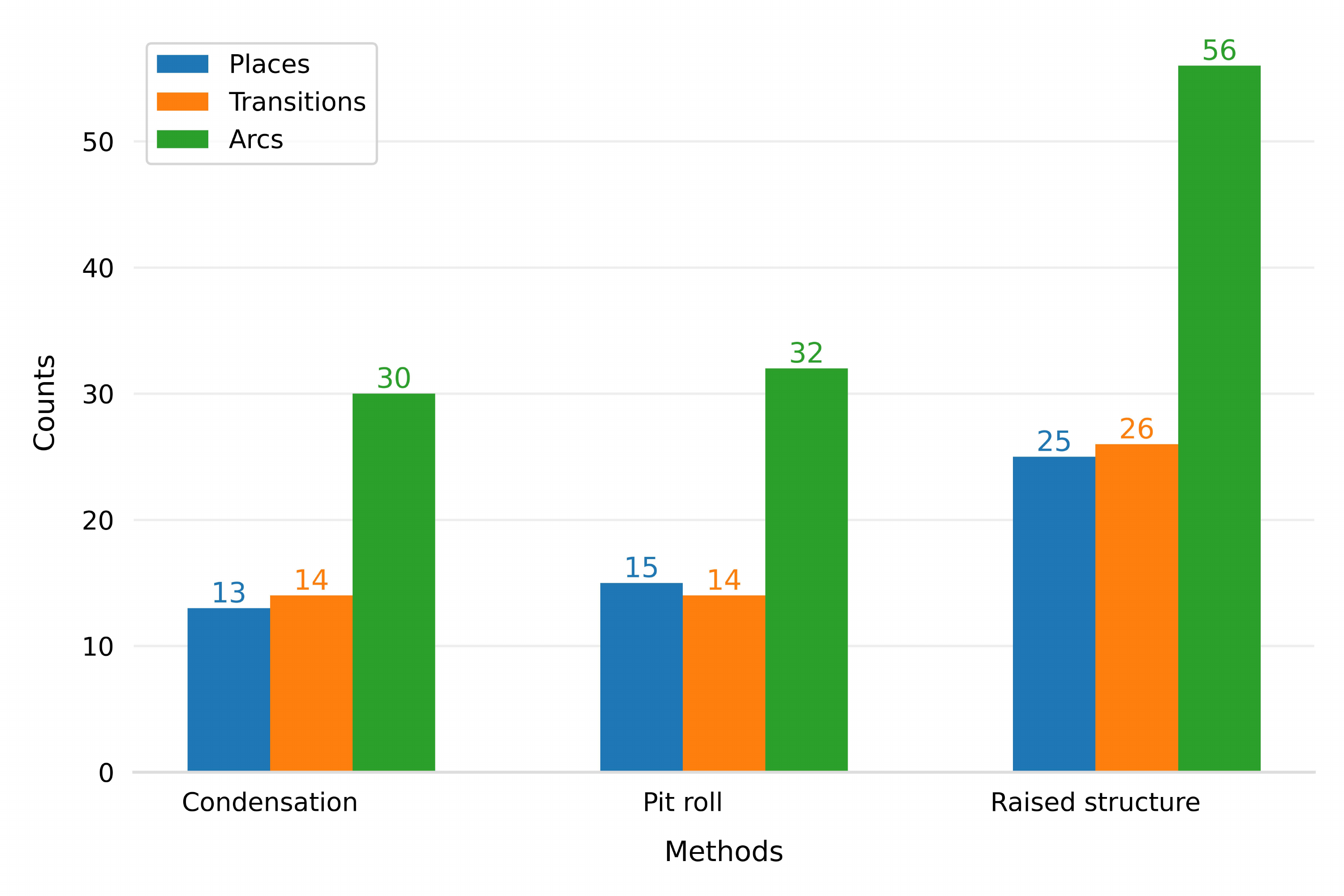}
	\caption{Number of Petri net elements contained in each tar production model}
	\label{fig:fig3}
\end{figure}

\subsection{The condensation method relies heavily on working memory capacity}

The condensation model scored the highest in the density metric (value = 0.082), followed by the pit roll model (value = 0.076), and then the raised structure model (value = 0.043; Figure 4A). The density value of the condensation model can be explained by the information processing peak during the use of fire. Considering the number of possible connections between transitions and places, and the size of the model, the density metric shows that this peak is more demanding than those in the other models. The use of fire is modelled with the transition ‘Light bark’, and the place ‘p6’. These elements are connected with other transitions and places of the model by three and four arcs, respectively (Figure 4A). The arc density is also higher during and after using fire in the condensation method than in the pit roll and raised structure methods. Nine arcs are located before the transition ‘Light bark’, and 21 arcs occur after in the condensation model.

\begin{figure}
	\centering
	\includegraphics[width=0.95\linewidth]{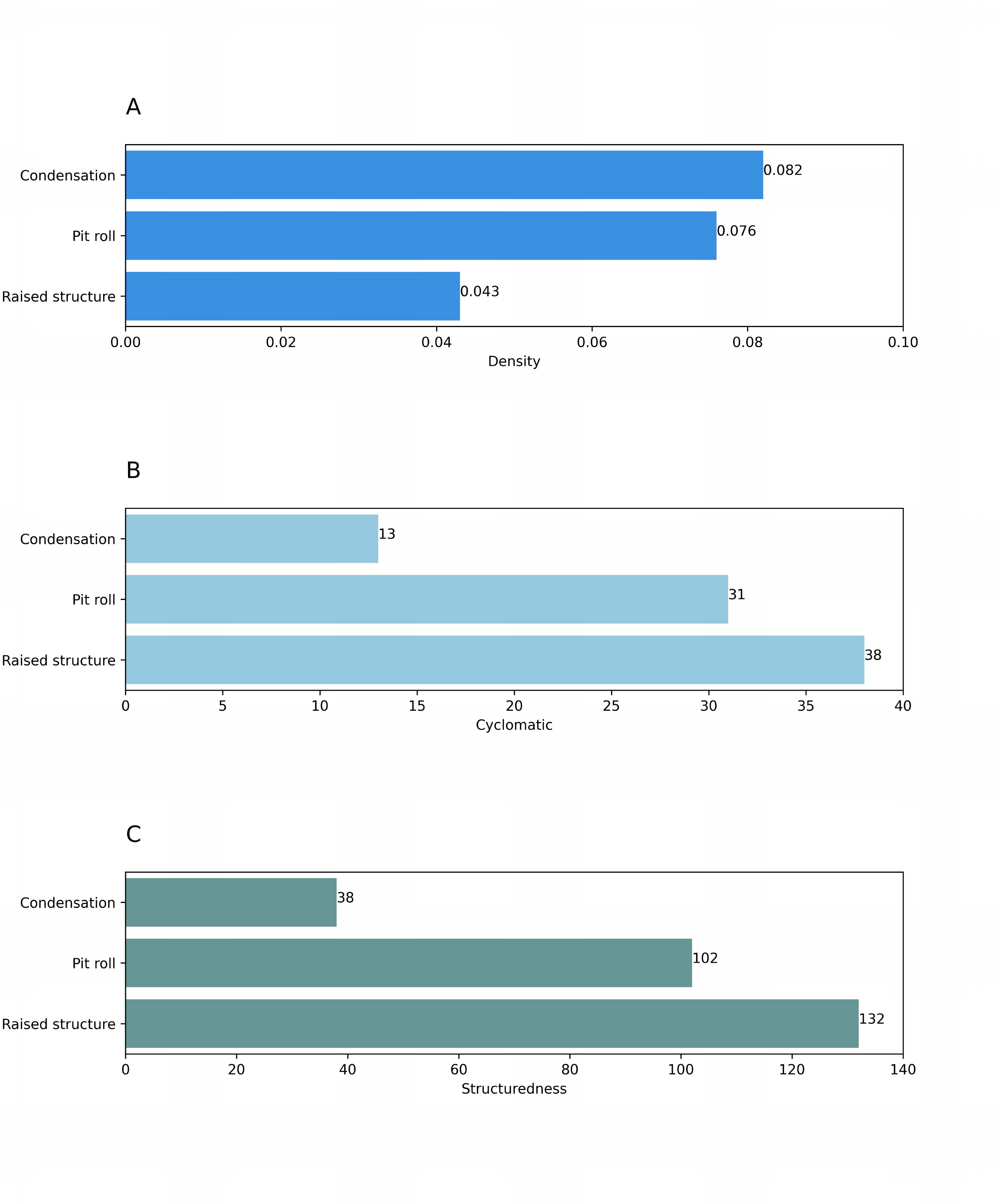}
	\caption{Complexity metrics values for the models of tar production. Density (a), Extended Cyclomatic (b) and Structuredness (c).}
	\label{fig:fig4}
\end{figure}

The pit roll model scored the second highest density of the three models (value= 0.075). When compared with the condensation model, this value results from a larger number of conditions to be fulfilled without a strong increment in the arc density. The pit roll model included only two more arcs and two more places than the condensation model. Three transitions and three places are connected each with more than two arcs (Figure 4C). Nineteen arcs in the pit roll model are located before the start of use of fire, represented by the transition ‘Place embers’, and the other 13 arcs appear after. This net structure shows that the arc density and the number of conditions are higher in preparations before using fire than during the use of fire.

The raised structure model showed the lowest density (value =0.043). There are four places and three transitions in the Petri net connected with more than two arcs with other elements (Figure 4C). However, the raised structure model almost doubles the number of places, transitions, and arcs compared with the other two models. This reduced the weight that the multiple arcs have in the density. When comparing net structures of the raised structure and condensation models, the raised structure model shows arcs that are more uniformly distributed between the preparations before using fire and the actions involved in the use of fire. Twenty seven arcs are located before the use of fire, represented by the transition ‘Light dome’, and 29 arcs appear after. The raised structure model is therefore balanced in terms of the actions associated with the preparations before using fire and the actions during and after the use of fire. 

\subsection{The pit roll and raised structured methods require control actions to reduce errors}

The condensation model scored the lowest in the extended cyclomatic metric (value = 13), followed by the pit roll model (value = 31), and then the raised structure model (value = 38, Figure 4B). A comparison of the number (Figure 6) and distribution (Figure 5) of directed edges, vertices and strongly connected components in the reachability graphs provide insights into the factors generating the differences in this metric.

\begin{figure}
	\centering
	\includegraphics{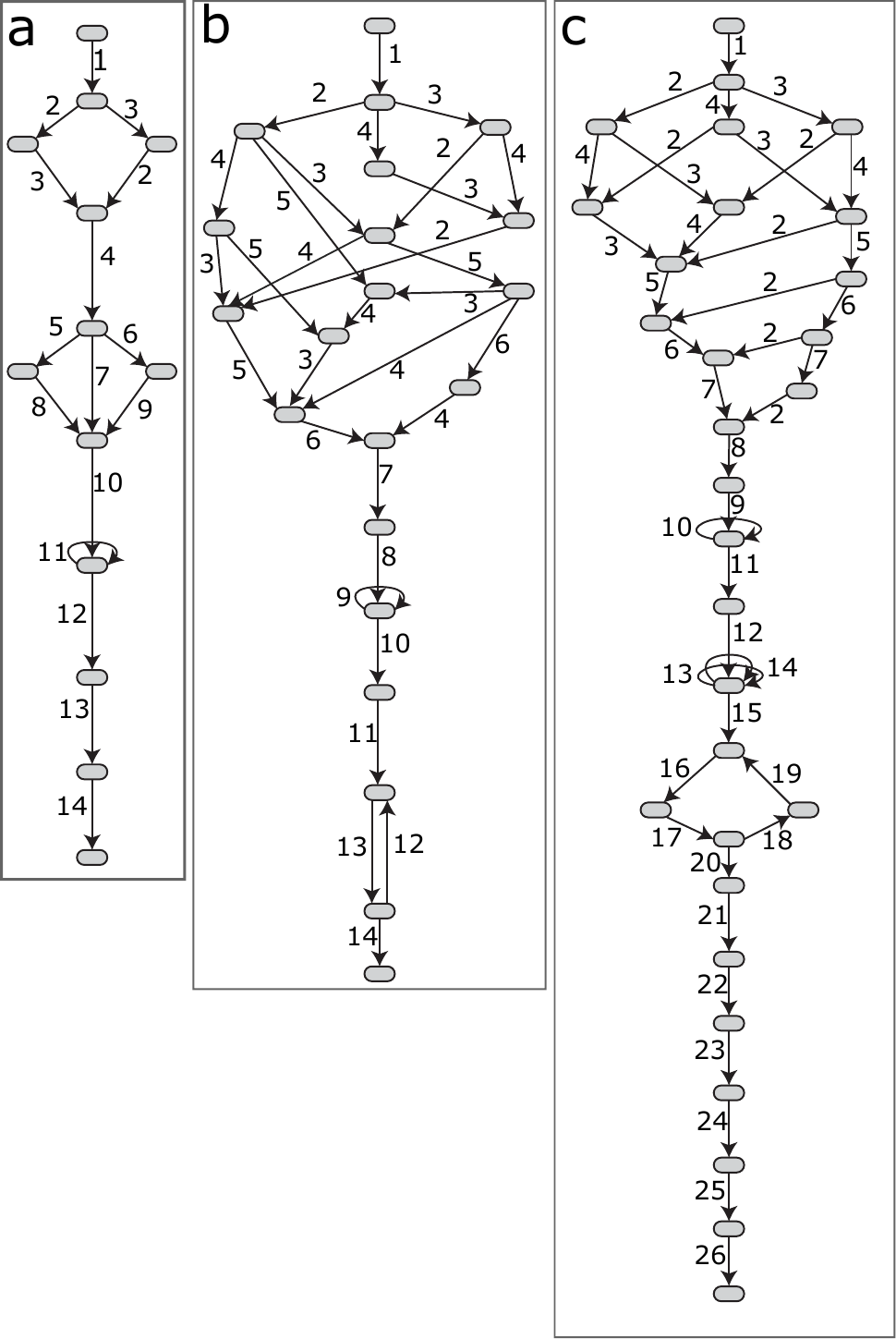}
	\caption{Reachability graphs showing states (vertices) and transitions (numbered edges) for each tar production model. (a) Condensation (1. Start, 2. Tear bark, 3. Place rock, 4. Light bark, 5. Grab, 6. Reignite, 7. Place lit bark, 8. Bark moves, 9. Bark extinguishes, 10. Start condense, 11. Hold bark, 12. Stop condense, 13. Scrape, 14. Store); (b) pit roll (1. Start, 2. Clean soil, 3. Make cup, 4. Make roll, 5. Dig, 6. Place cup, 7. Place roll, 8. Place embers, 9. Fan embers, 10. Cool pit, 11. Dig roll \& cup, 12. Reheat roll, 13. Collect tar, 14. Store); (c) raised structure (1. Start, 2. Make roll, 3. Dig pit, 4. Make cup, 5. Place cup, 6. Place net, 7. Place pebbles, 8. Place roll, 9. Make dome, 10. Fix dome, 11. Place firewood, 12. Light dome, 13. Add firewood, 14. Fix dome2, 15. Fire stops, 16. Cool dome, 17. Open dome, 18. Dome fumes, 19. Close dome, 20. Fuming stops, 21. Remove dome, 22. Remove roll, 23. Remove net \& pebbles, 24. Remove cup, 25. Collect tar, 26. Store).}
	\label{fig:fig5}
\end{figure}

\begin{figure}
	\centering
	\includegraphics[width=0.8\linewidth]{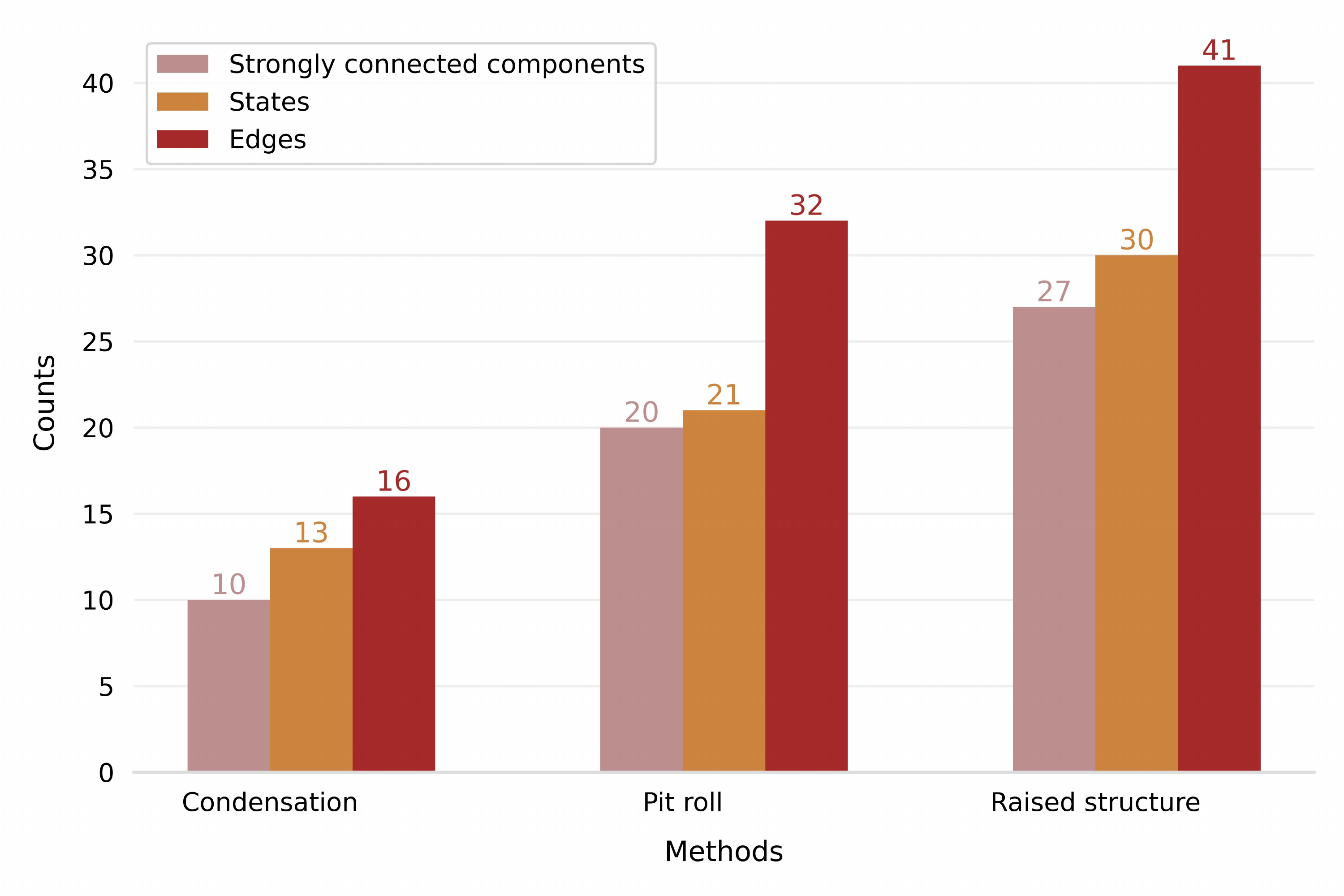}
	\caption{Number of elements in the reachability graphs of the tar production models}
	\label{fig:fig6}
\end{figure}

The low score of the condensation model (value = 13), and the low number of reachable states are explained by the number of transitions that can be executed concurrently and the small number of sequential actions. Thirty eight percent (N=5) of the total reachable states occur during the preparations before the use of fire. The only two transitions (‘Tear bark’ and ‘Place rock’) that may occur concurrently at any given marking are part of the preparations. The potential for concurrent actions in the condensation model is the lowest of the three models. The largest strongly connected components in the condensation model has four reachable states representing 30\% of the total. This is the highest percentage for any strongly connected component in the three models. The largest strongly connected component in the condensation model occurs after the transition ‘Light bark’ (Figure 5A). One repetitive action (‘Hold bark’) occurs as a self-loop during the use of fire. 
The pit roll model scored the second highest for the extended cyclomatic metric (value=31). Compared with the condensation model, the reachability graph of the pit roll model has almost double the number of reachable states, and twice the number of edges and strongly connected components (Figure 5B; Figure 6). Seventy six percent (N=16) of the total reachable states occur during the preparations for the use of fire, where a maximum of three transitions in any combination from the set ‘Make cup’; ‘Clean soil; ‘Dig pit’; ‘Place cup’ and ‘Make roll’ can occur concurrently (Figure 2B). The maximum number of possible concurrent transitions in the pit roll suggests that concurrent actions are more important in the pit roll method than in the condensation method. The largest strongly connected component in the reachability graph of the pit roll model has two reachable states and represents 9.5\% of the total. One repetitive action (‘Fan’) occurs as a self-loop during the use of fire.

The raised structure model scored the highest for the extended cyclomatic metric (value= 38). The raised structure model is the longest of the three models and also shows the largest number of sequences (Figure 5C; Figure 6). This model also includes transitions with the potential to be executed concurrently before the use of fire, and repetition of actions during the use of fire. Fifty three percent (N=18) of the total reachable states occur during the preparations for the use of fire. A maximum of three of the transitions ‘Make cup’, ‘Place net’, ‘Place pebbles’, ‘Dig pit’, ‘Place cup’, and ‘Make roll’ can occur concurrently in any combination (Figure 2C). The largest strongly connected component in the reachability graph of the raised structure model has four reachable states and represents 13\% of the total. This strongly connected component, generated by a cycle of actions during the opening of the structure, is twice the size of that from the pit roll and has the same number of reachable states as the condensation model. Three other actions (‘Fix dome’; ‘Add firewood’; ‘Fix dome 2’) are repeated as self-loops, separated by sequences of actions during the use of fire. 

\subsection{The raised structure contains more embedded information than the condensation or pit roll methods}

The condensation model scored the lowest in the structuredness metric (value = 38), followed by the pit roll (value = 102), and then the raised structure (value = 132; Fig 3C). The metric values indicate that the raised structure model requires more planning and acquiring more knowledge about the production process than the pit roll and condensation models do. All three Petri net models showed sequences, whiles, marked graphs, and state machines (Figure 7), organized in a three tier hierarchical structure. Sequences are found in the deepest tier inside the marked graph of the pit roll and raised structure models, but they are also found embedded in the state machine components in the second tier in all nets. Sequences are the most dominant component representing 68\% of all components matched. The raised structure model shows the highest number of sequences (N = 12) and the pit roll model the lowest (N = 3). 

\begin{figure}
	\centering
	\includegraphics[width=0.8\linewidth]{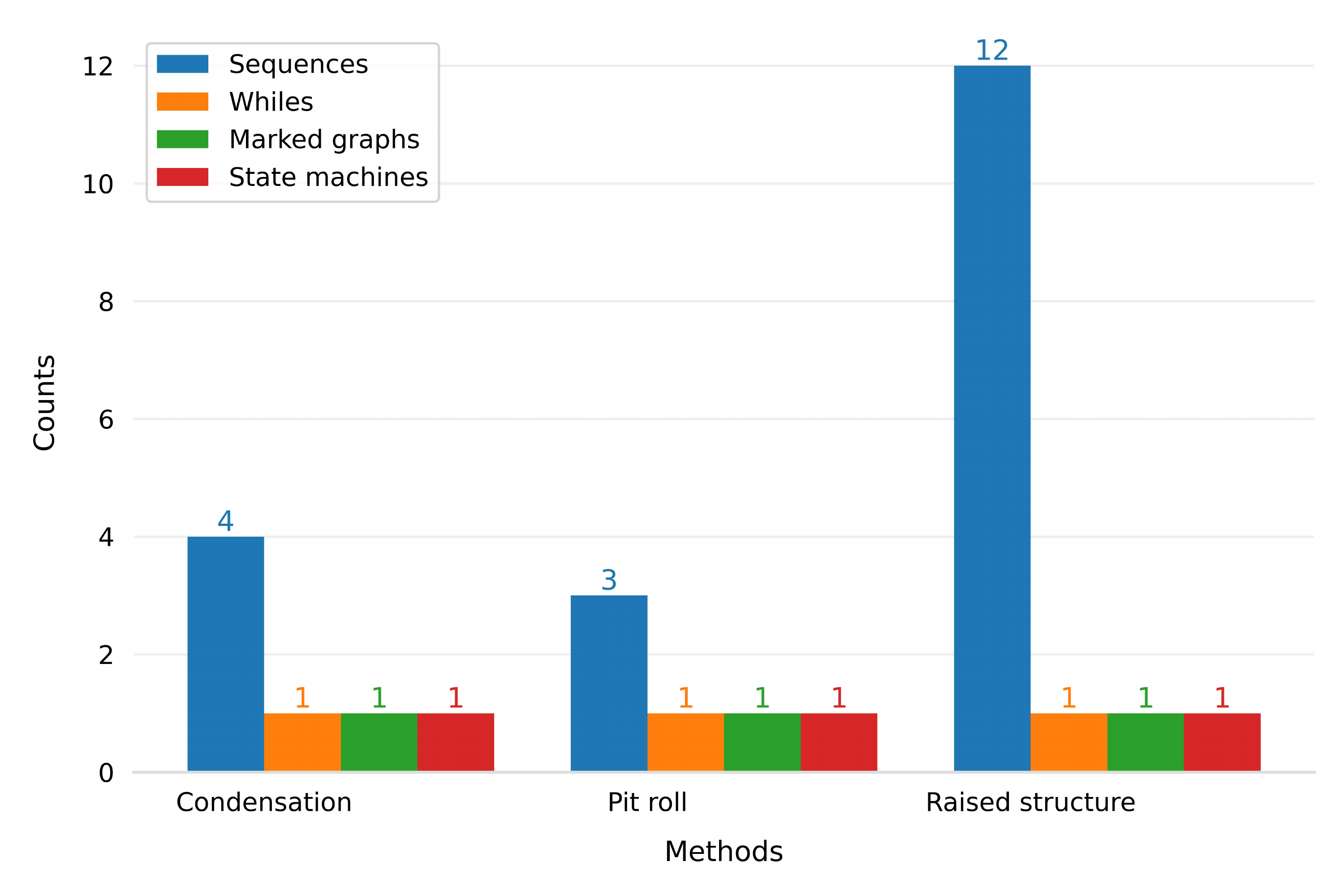}
	\caption{Number of constructs found with the structuredness metric algorithm in the condensation, pit roll and raised structure models}
	\label{fig:fig7}
\end{figure}

The while components appear in the second tier of the deconstructed nets, embedded within the state machine components. Each model contained one while component. 
The marked graph components were found with different sizes in the second tier of the deconstructed nets. The marked graphs relate to the preparations before the use of fire. The marked graph of the condensation model is the smallest, collapsing four places and four transitions. The marked graph of the pit roll model collapses eight places and seven transitions, and has one embedded sequence. The raised structure model has a marked graph that collapses nine places and has two embedded sequences.

The top tier of the hierarchical structure of the decomposed nets is a state machine component. The sequences, whiles and marked graphs are embedded in this tier. The top tier shows different sizes for each model with macro transitions representing the embedded sequences, whiles, and marked graphs described above. At the top tier of the deconstructed net, the condensation model had a total of four places, four macro transitions, and one transition. The pit roll model showed four places, one macro transition, and three transitions. Finally, the raised structure model showed five places, four macro transitions, and three transitions. 

\section{Discussion}

The experiments, the Petri net models, and the metrics show that the condensation method relies on working memory use. The complexity metrics also show that the raised structure method relies on cognitive functions that combine the use of different cognitive processes, such as working memory, planning, and learning.
Based on the structure of the Petri net models and the density metric, the condensation method imposes the most intense cognitive load in working memory of all three methods (Figure 6). The values of the density metric show that actions and resources are more interconnected in the condensation model. Having more actions and resources interacting at the same time requires accessing more information at a given time. The cognitive load in the condensation method is generated by the attention required to maintain the pieces of lit bark before and during the tar condensing on the cobble. The results for the pit roll model indicate that it requires less attention than the condensation model and that the most interconnected elements and actions occur during the preparations before using fire. Working memory is less intensely used in the pit roll and raised structure methods because peaks in information processing are smoother compared with the condensation model. In the raised structure method, actions are less interconnected and more distributed through the entire process, so the focus of the working memory resources can be on fewer actions at a time. Working memory and allocation of attention are two modern human cognitive resources that allow us to process and navigate through large amounts of information (\cite{liederResourcerationalAnalysisUnderstanding2020}). 

The lower likelihood of errors shown in the cyclomatic metric suggests that the condensation method is a simpler solution because fewer possible paths exist to obtain the tar product. This means that the variability in the way the process works, and thus the underlying behavior is less complex than in the pit roll and raised structure methods. The pit roll model is similar to the raised structure in that it shows more possible paths and a higher likelihood of errors than the condensation model. The potential for concurrency generates most of the behavioral complexity of the pit roll method. The results show that the possible paths to reach the final product and the possibility of errors of the raised structure model are generated by combinations of concurrent activities, actions executed as cycles, and repetitive actions executed as self-loops. The raised structure has the largest number of reachable states, paths, and possible errors to reach the final product. 

The condensation method is the only method where fewer reachable states are present before fire than after (Figure 2 and Figure 5). The results show that the behavioral complexity in the condensation model after the use of fire is associated with repetitive actions that may occur as cycles or self-loops. The potential for concurrency and the largest number of reachable states of the pit roll and raised structure methods occurs before fire, suggesting that the technical behavior during preparation is more cognitively demanding. The use of fire is unique among actions in tar production. No possible concurrent actions occur after the use of fire in any of the three models, meaning that fire use is a synchronization event where all tasks that can be conducted concurrently come together, and attention shifts from being divided among multiple possible actions, to being focused on one action at a time. This is done to finish the process and obtain the product. The need for synchronization of material flows obtained via concurrent events is a feature that appear in human made systems, for example, in ceramic production systems that mix clays to obtain pottery with specific properties (\cite{costinUseEthnoarchaeologyArchaeological2000}).

The models and the structuredness metric suggest that the condensation method is easier to learn than the two other methods. The embedded components of the condensation model, and especially its marked graph, are smaller, suggesting that the amount of information required to execute the process is also smaller. The condensation model scored the lowest in the structuredness metric, despite having more transitions in its state machine component than the pit roll model. The marked graph of the pit roll model is the second largest and has more actions with potential for concurrent activities than the condensation model, making the score of the pit roll model the second highest in the structuredness metric. The raised structure model scored the highest in the extended cyclomatic metric because it has more than twice the number of sequences in the workflow and its marked graph and state machine components are larger than the other two models. This suggests that the raised structure method requires more effort to understand the elements in the production process because it shows a more elaborated and larger process structure than the pit roll and condensation models. The results also indicate that the pit roll and raised structure model have at least three times more embedded information in their process structures than the condensation model. The raised structure model shows the largest amount of information in its structure, meaning that greater understanding and planning is needed to complete the process. The experiment observations further support this and showed that the pit roll and the raised structure methods required more planning for their execution. The differences in understandability of the three methods show that reasoning and planning, common cognitive processes used by modern humans (\cite{callawayRationalUseCognitive2022, liederResourcerationalAnalysisUnderstanding2020}), may have been involved in some of the tar production methods available to Neanderthals. 

Currently, we do not know what production methods Neanderthals used. Recent studies show evidence of other production processes like plant cooking (\cite{kabukcuCookingCavesPalaeolithic2022}) and fire use (\cite{macdonaldMiddlePleistoceneFire2021}) with multiple steps and components. This evidence of complex behaviors supports the possibility that the 50ka year old Dutch Zandmotor tar could have been produced with methods similar to the raised structure method (\cite{niekusMiddlePaleolithicComplex2019}). We show that irrespective of the methods being used, prehistoric tar making may have required aspects of cognition analogous to that of contemporary modern humans. The results of our study lend further support to the hypothesis that Neanderthals and modern humans may have had similar working memory capacities and likely employed them in comparable ways (\cite{ambroseCoevolutionCompositetoolTechnology2010, haidleWorkingMemoryCapacity2010}; but see also (\cite{kochiyamaReconstructingNeanderthalBrain2018, wynnExpertNeandertalMind2004}). This is observed in the number of features that must be kept in working memory in each of the three methods. The working memory capacity for contemporary humans is estimated at around four items (\cite{cowanMagicalNumberShortterm2001}), but recent studies show that working memory usage maintains two to three features simultaneously (\cite{draschkowWhenNaturalBehavior2021}). A peak of three retained features in memory occurs in the condensation method. This is related to three conditions: the location of the bark against the rock, the state of the lit bark, and the moment when condensation starts. The attention required to monitor these three conditions creates the relations between resources and activities that gives the condensation method its high-density metric value. If the production processes were executed in one event without interruption, then a maximum of two concurrent activities in the condensation method, and three activities in the pit roll and raised structure methods require the makers to store these activities in their working memory to avoid repetition. This is also required for components of composite tools that are produced asynchronously (\cite{hoffeckerComplexityNeanderthalTechnology2018, hoffeckerStructuralFunctionalComplexity2018}). We suggest that the working memory requirements for ancient tar technology were comparable to the use of working memory by modern humans today. These results are consistent with evidence from other Neanderthal cognitively complex behaviors such as deep cave activity (\cite{jaubertEarlyNeanderthalConstructions2016}), cave painting (\cite{hoffmannUThDatingCarbonate2018}), use of jewelry and body painting (\cite{bednarikBeadsPendantsPleistocene2001}), and deliberate burial (\cite{renduEvidenceSupportingIntentional2014}) identified over the last decades.
The three production processes all contain choices that require inhibitory control. To ensure that the process will produce tar when its execution terminates, makers are required to control their need of obtaining tar. All the models ensure termination of the production processes and the cyclomatic metric evaluates every possible combination of actions in modeled production processes. However, in real world situations, we cannot ensure that a production process will yield the desired product if it terminates because not all events will occur successfully every time. Interference, random events, or urgency in obtaining the product may prevent makers from fetching tar, even if all steps in the process are executed. For example, opening the dome of the raised structure without giving enough time to reach high temperatures, increase the possibility of not obtaining tar even if all the steps in the process are executed. Therefore, self-control is required in tar production. This type of self-control in the production of material culture is argued to date back to the Early Stone Age and a prerequisite for any tool-making and extended problem solution distance behaviors (\cite{haidleWorkingMemoryCapacity2010, kohlerMentalityApes1925, lombardCognitionCapuchinRock2019}). For example, in stone-tool making (\cite{pargeterUnderstandingStoneToolmaking2019}) and specifically in the production of symmetrical Acheulean handaxes (\cite{greenNotJustVirtue2020}) self-control is required to invest time to acquire the required stone knapping skills. The modelling approach and the metrics presented here can be helpful to test such hypotheses in the future.

If the information required to produce a technology is acquired cumulatively (\cite{caldwellUsingExperimentalResearch2020, wadleyWhatStimulatedRapid2021}; but see also  \cite{vaesenHumanCultureCumulative2021}), processes with low understandability (i.e., more information), such as the raised structure, are likely to emerge later in the development of a technology. Conversely, processes with high understandability are likely to be more prevalent during the emergence of a technology. In this light, the condensation method could have been discovered first (\cite{schmidtBirchTarProduction2019}). It relies on materials directly available in the environment and the process of tar formation can be directly observed in an open fire. This technique may even qualify as a latent solution (\cite{schmidtBirchTarProduction2019, tennieEarlyStoneTools2017, tennieIslandTestCumulative2016}; presented under the right circumstances to an individual and not requiring teaching. For Neanderthal tar making via condensation, these circumstances must have included sufficient working memory, access to birch bark, a suitable rock, a tool for scraping, and fire. The other production methods have more embedded information, making them more difficult to learn. They are, therefore, unlikely to be latent solutions. It is more likely that, if used, the technological know-how of the pit roll and raised structure techniques were transmitted culturally (\cite{tomaselloCulturalLearning1993}). These two production methods also rely on a greater planning depth and inhibition ability, and the integration of working memory with other cognitive processes. 
Our study and method are not without limitations. Here we studied a single technology using three possible production methods and in reality, Neanderthals may have used more or different methods (\cite{kochNewMethodBirch2022, kozowykExperimentalMethodsPalaeolithic2017, pomstraProductionBirchPitch2010}). The results are by no means a complete representation of the complexity of the Neanderthal technological world; this would require modelling of the production of bone, wood, stone tools, and other technologies like fire (\cite{adlerEarlyLevalloisTechnology2014, arangurenWoodenToolsFire2018, leder51000yearoldEngraved2021, schlangerUnderstandingLevalloisLithic1996, sorensenNeandertalFiremakingTechnology2018, soressiNeandertalsMadeFirst2013}). In future works, all the possible production methods of multiple technologies from ancient technological systems should be compared to illuminate the trends in technological and cognitive solutions available in the past. In addition, some of the tar production techniques are scalable and we did not model for that here. The models and metrics have limitations to study the scalability of the production process or the effects of resource availability in the process complexity. Models scaling up the production will influence the results of the metrics because the number of connections between actions and resources, and amount of information embedded in the structure of the models will be dependent on how much the processes are scaled up. Other classes of Petri nets such as place/transitions nets are more adequate to explore these problems (\cite{fajardoModellingMeasuringComplexity2022}). In addition, with Petri nets we model a reconstructed version of the past; missing details in this reconstruction may influence measured outcomes. For some technologies, like refitted lithic technologies much data may be present, whereas for others, like basketwork, there is heavy reliance on experimental archaeology. The used metrics are designed for modern human cognition and one can question their relevance to ancient cognition. In this study we work from the assumption that both Neanderthals and modern humans share evolutionary traits, and we considered the metrics suitable. 
Neanderthals are increasingly argued to have technological and cognitive capabilities comparable to modern humans (\cite{macdonaldMiddlePleistoceneFire2021, pitarchmartiSymbolicRoleUnderground2021, roebroeksNeandertalsRevised2016, villaNeandertalDemiseArchaeological2014}). We have shown it is possible to derive the cognitive requirements of technologies, making such arguments concrete and measurable. Petri nets and their derived measures are unique because they allow us to also compare complexity across different materials and technologies. Petri nets are also a promising tool to study older technologies where inferences are harder to make compared to the Middle Paleolithic, shedding light on the cognitive requirements of our earliest tool making ancestors.

\section{Conclusion}

The method presented here links process complexity of ancient technologies to cognitive processes. At face value the condensation method relies on working memory. The other two methods rely much less on working memory, but are more demanding in terms of knowledge density, and require cultural learning, self-control, and planning. The Petri net approach to measure complexity presented here has proven to be useful for the systematic analysis and comparison of ancient technologies and their production systems. Independent of which proposed tar production methods were used in the past, the Petri net models and complexity metrics suggest that Neanderthals probably relied on several cognitive traits that archaeologists often associate with modern behavior. Our multi-angled approach to the embedded information in technological behaviors does justice to the different and idiosyncratic forms of technological complexity and cognitive traits hominins may have had. Future studies can extend this implementation to other technological behaviors to improve our understanding of human and technological evolution.

\section{Acknowledgments}

We thank Alessandro Aleo for his collaboration conducting experiments. We also thank to the educational archaeological site Masamuda in Vlaardingen (the Netherlands) for their generous use of space for the experiments. This research was supported as part of the Ancient Adhesives project, funded by the European Research Council (https://erc.europa.eu/) under the European Union’s Horizon 2020 research and innovation programme grant agreement No. 804151 (grant holder GHJL). The funders had no role in study design, data collection and analysis, decision to publish, or preparation of the manuscript.

\bibliographystyle{rusnat}
\bibliography{references}  



\end{document}